\documentclass[11pt]{article} 
\newcommand{\be}{\begin{equation}}
\newcommand{\ee}{\end{equation}}
\newcommand{\bea}{\begin{eqnarray}}
\newcommand{\eea}{\end{eqnarray}}
\newcommand{\sn}{{\rm sn}}

\newcommand{\dn}{{\rm dn}}
\newcommand{\cn}{{\rm cn}}
\newcommand{\sech}{{\rm sech}}

\begin{document}
\vspace{.5in} 
\begin{center} 
{\LARGE{\bf Connections Between Complex PT-Invariant Solutions  
and Complex Periodic Solutions of Several Nonlinear Equations}}
\end{center} 

\vspace{.3in}
\begin{center} 
{\LARGE{\bf Avinash Khare}} \\ 
{Physics Department, Savitribai Phule Pune University, 
Pune, India 411007}
\end{center} 

\begin{center} 
{\LARGE{\bf Avadh Saxena}} \\ 
{Theoretical Division and Center for Nonlinear Studies, Los
Alamos National Laboratory, Los Alamos, NM 87545, USA}
\end{center} 

\vspace{.9in}
{\bf {Abstract:}}  

We point out novel connections between complex PT-invariant solutions of
several nonlinear equations such as $\phi^4$, $\phi^6$, sine-Gordon, hyperbolic 
sine-Gordon, double sine-Gordon, double hyperbolic sine-Gordon, mKdV, etc.  
We then use these connections to obtain several new complex PT-invariant 
periodic solutions of some of these equations.

\newpage 
  
\section{Introduction}

During the last few years parity-time inversion or PT symmetry \cite{ben} has 
received a lot of attention mainly because of its possible applications in open 
systems with balanced loss and gain, particularly in photonics  \cite{sch,ben1,pen,rut}. 
Recently, we have obtained PT-invariant kink as well as periodic kink solutions \cite{ks1,ks2}  
of several models such as $\phi^4, \phi^6$, sine-Gordon (SG), double sine-Gordon 
(DSG) and double sine hyperbolic-Gordon (DSHG) equations of the form $\tanh(x) \pm 
i\sech(x)$ and $\sn(x, m) \pm i \cn(x, m)$ as well as $\sn(x, m) \pm i \dn(x, m)$, 
respectively.  It is worth recalling that in all these cases, one had a kink solution  
of the form $\tanh(x)$ and a periodic kink solution of the form $\sn(x, m)$. 
Here $\sn(x, m), \cn(x, m)$ and $\dn(x, m)$ are Jacobi elliptic functions
(JEF) with $m$ being the modulus of the JEF ($0 \le m \le 1$) \cite{as}. 
Soon afterward, it was pointed out by Cen and Fring \cite{fri} that since these 
models are translationally invariant, if $\tanh(x)$ is a solution, then so must be 
$\tanh(x+a)$. On taking $a = \pm i\pi/4$, one immediately obtains the complex 
PT-invariant kink solution $\tanh(x) \pm i\sech(x)$ mentioned above with half the 
width and the same amplitude as compared to the $\tanh(x)$ solution thus proving that 
whenever a model admits a kink solution of the form $\tanh(x)$, the same model will 
necessarily also admit a PT-invariant complex kink solution of the form $\tanh(x) \pm 
i\sech(x)$. 

The important question is whether a similar argument also holds good for the 
corresponding periodic kink solutions in terms of the JEF functions. In 
particular, by 
translational invariance, if some model has a periodic solution of the form $\sn(x, m)$, 
then presumably it also has a solution of the form $\sn(x+a, m)$. If we now take $a = iK'(m)/2$ 
(note that at $m = 1$, $iK'(m)/2 = i\pi/4$, where $K'(m)$ is the complete elliptic integral 
of the first kind with complementary modulus $m'=\sqrt{1-m^2}$) then on using the addition 
formula for $\sn(x, m)$
\be\label{sna}
\sn(a+b, m) = \frac{\sn(a, m)\, \cn(b, m)\, \dn(b, m) +\sn(b, m)\, \cn(a, m)\, 
\dn(a, m)}{1-m\sn^2(a, m)\, \sn^2(b, m)}\,, 
\ee
we find that
\be\label{I1}
\sn[u \pm iK'(m)/2, m] = \frac{(1+\sqrt{m})\sn(u, m) \pm i \cn(u, m)\,
\dn(u, m)}{m^{1/4} [1+\sqrt{m} \sn^2(u, m)]}\,.
\ee
Here we have used the fact that \cite{as} 
\be\label{snb}
\sn[iK'(m)/2, m] = \frac{i}{m^{1/4}}\,,~~\cn[iK'(m), m] 
= \frac{(1+\sqrt{m})^{1/2}}{m^{1/4}}\,,~~\dn[iK'(m), m] = 
(1+\sqrt{m})^{1/2}\,. 
\ee
Unfortunately, solution (\ref{I1}) is very different from the well known 
complex periodic kink solutions \cite{ks1,ks2} of the form 
$\sn(x, m) \pm i \cn(x, m)$
or $\sn(x, m) \pm i \dn(x, m)$ and there is no special case under which it
goes to either of these complex PT-invariant periodic kink solutions. 

Inspired by this, we have explored new solutions and have been able to obtain 
even more general complex PT-invariant periodic kink solutions of several 
nonlinear equations. Further, by extending similar arguments, we have also 
obtained even more general complex PT-invariant periodic 
pulse solutions of several nonlinear equations. One of the purposes of this 
paper is to discuss in detail some of these solutions.

The other purpose of this paper is to point out novel connections between 
complex solutions of either different nonlinear models or even the same
model but with different values of the parameters. As an illustration,
consider the celebrated $\phi^4$ static field equation 
\be\label{I2}
\phi_{xx} = a \phi + b \phi^3\,.
\ee
Now observe that under $\phi \rightarrow \pm i\phi$, this equation goes over
to the $\phi^4$ field equation
\be\label{I3}
\phi_{xx} = a \phi - b \phi^3\,.
\ee
It is then clear that if for a given $a, b$, $\phi^4$ field Eq. (\ref{I2}) admits 
a complex solution $\phi_1$, then the same model will also admit  
the complex solution $\pm i\phi_1$ for the same $a$ but opposite $b$. 
We discuss similar connections between several models, e.g., sine-Gordon (SG) 
\cite{dj,asegur} and sine hyperbolic-Gordon (SHG) \cite{shg}, double sine-Gordon (DSG) 
\cite {dsg} and double sine hyperbolic-Gordon (DSHG) \cite{razavi,khs,hks},  
both focusing and defocusing mKdV \cite{dj,asegur} and also in the $\phi^{4n+2}$ - 
$\phi^{2n+2}$ - $\phi^2$ models where $n = 1,2,3,...$. Using such connections 
we then obtain several novel complex solutions of some of these equations.

The plan of the paper is as follows. In Sec. II we discuss novel connections
between the complex solutions of several real nonlinear equations. Using
these connections, we obtain several complex PT-invariant periodic solutions
of the celebrated $\phi^4$ equation. In Sec. III we use the connection 
between the complex solutions of the focusing and the defocusing mKdV 
equations and obtain several complex PT-invariant periodic solutions of these 
two equations. In Sec. IV, using the connection between the complex 
PT-invariant 
solutions of the DSG and the DSHG equations we obtain several new periodic kink 
and pulse solutions of both the models. In Sec. V we discuss a coupled $\phi^4$ 
model discussed by us before \cite{ks3} and obtain a large number of new solutions 
of this coupled model. Finally, in Sec. VI we summarize the main results obtained in  
this paper and point out some of the open problems.

\section{Superposed Complex PT-invariant Periodic solutions of $\phi^4$}

Let us consider the celebrated static $\phi^4$ field equation (\ref{I2}).
We will first show that apart from the two well known complex PT-invariant
periodic kink solutions, it admits yet another (and in fact more general)
complex PT-invariant periodic kink solution.
 
\subsection{Superposed Complex PT-Invariant Periodic Kink Solutions}

It is well known \cite{phi4} that the periodic kink solution of $\phi^4$ field Eq. (\ref{I2}) is 
\be\label{2.2}
\phi = A \sn(\beta x, m)\,,
\ee
provided
\be\label{2.3}
b A^2 =  2\beta^2\,,~~a= -(1+m) \beta^2\,.
\ee
Note that such a solution exists provided $b > 0, a < 0$. In the limit $m = 1$
we then have the celebrated kink solution
\be\label{2.2a}
\phi = A \tanh(\beta x)\,,
\ee
provided
\be\label{2.3a}
b A^2 =  2\beta^2\,,~~a= -2 \beta^2\,.
\ee

What about the corresponding complex PT-invariant periodic solutions? It
is well known \cite{ks1,ks2} that the $\phi^4$ field Eq. (\ref{I2}) admits two such 
solutions. The first one is
\be\label{2.2b}
\phi = A[\sqrt{m} \sn(\beta x, m) \pm i \dn(\beta x, m)]\,,
\ee
provided
\be\label{2.3b}
a = -(2m-1)/2 \beta^2\,,~~b A^2 = \beta^2\,, 
\ee
and the second solution is
\be\label{2.2d}
\phi = A\sqrt{m} [\sn(\beta x, m) \pm i \cn(\beta x, m)]\,,
\ee
provided
\be\label{2.3d}
a = -(2-m)/2 \beta^2\,,~~b A^2 = \beta^2\,.
\ee

Are the solutions (\ref{2.2b}) and (\ref{2.2d}) the most general complex 
PT-invariant periodic kink solutions? We now show that the answer to the 
question is no.

Let us recall that $\phi^4$ model is a translation invariant system. Thus, if
$\sn(\beta x, m)$ is a solution then so is $\sn(\beta x \pm iK'(m)/2,m)$.
Inspired by the identity (\ref{I1}), we now inquire if there is a more general 
complex, PT-invariant periodic kink 
solution of the $\phi^4$ field Eq. (\ref{I2}).  In particular, we start with the 
ansatz
\be\label{2.4}
\phi = \frac{A\sn(\beta x, m) \pm iB \cn(\beta x, m)\, \dn(\beta x, m)}
{1+D\sn^2(\beta x,m)}\,,
\ee
where $A, B, D, \beta$ have to be determined in terms of the parameters $a, b$ 
of the model. Here $D > -1$ so as to avoid any singularity. After a lengthy 
algebra we find that indeed Eq. (\ref{2.4}) is 
a novel periodic kink solution of the $\phi^4$ Eq. (\ref{I2}) provided
\be\label{2.5}
a = -(1+m)\beta ^2 < 0\,,~~b A^2 = 2(1+D)(m+D)\beta^2\,,~~bB^2 = 2D\beta^2\,.
\ee
It follows from here that either $D > 0$ or $-1 < D < -m$.
Further, note that while $a < 0$, $b > 0$ or $b < 0$ depending on  if
$D > 0$ or $ -1 < D < -m$. 
Thus, unlike the periodic kink solution (\ref{2.3}),
this solution can exist even when $b < 0$. It is then clear 
that the complex periodic kink solution (\ref{2.4}) is a new periodic
kink solution which cannot be obtained by translation from the periodic kink
solution (\ref{2.2}) since that solution only exists if $b > 0$. Further,
it is also obvious that the new periodic kink solution (\ref{2.4}) is 
distinct from the two complex PT-invariant periodic kink solutions 
(\ref{2.2b}) and (\ref{2.2d}) since those two solutions also exist only if
$b > 0$. Thus we now have three distinct complex PT-invariant 
periodic kink solutions of the $\phi^4$ field Eq. (\ref{I2}). 

One obvious question is: are there even more complex PT-invariant periodic kink 
solutions of the $\phi^4$ field Eq. (\ref{I2})?  Another question is: if at least
in the case $D> 0$ and hence  $b > 0$, does the solution (\ref{2.4}) simply 
follow from the real periodic kink solution (\ref{2.2}) by a translation? We now 
show that this is indeed the case. 

As argued above, since $\phi^4$ model is a translation invariant system, hence if
$\hat{A}\sn(\beta x, m)$ is a solution then so is 
$\hat{A}\sn(\beta x+ia, m)$. On using the addition theorem for $\sn(x,m)$ as 
given by Eq. (\ref{sna}) and the fact that \cite{as} 
\be\label{2.5d}
\sn(ia, m) = i\frac{\sn(a, 1-m)}{\cn(a, 1-m)}\,,~~\cn(ia, m) = \frac{1}
{\cn(a, 1-m)}\,,~~\dn(ia, m) = \frac{\dn(a, 1-m)}{\cn(a, 1-m)}\,, 
\ee
it is easily shown that the corresponding solution has precisely the form of 
the solution (\ref{2.4}) with
\be\label{2.5f}
A = \hat{A} \frac{\dn(a, 1-m)}{\cn^2(a, 1-m)}\,,~~B = \hat{A}
\frac{\sn(a, 1-m)}{\cn(a, 1-m)}\,,~~D = \frac{m \sn^2(a, 1-m)}{\cn^2(a, 1-m)}
> 0\,,
\ee
and the condition (\ref{2.5}) takes the form 
\be\label{2.5g}
a = -(1+m) \beta^2 < 0\,,~~b \hat{A}^2 = 2m \beta^2\,,
\ee
which is precisely the condition under which the solution 
$\hat{A} \sn(\beta x, m)$ holds good. It must be emphasized once again
that if instead one considers solution (\ref{2.4}) with $D < 0$, then such
a solution cannot follow from the translation invariance as considered above 
since, as is clear from Eq. (\ref{2.5f}), such a translation always gives $D > 0$.   
Summarizing, we thus have a new PT-invariant periodic kink solution
\be\label{2.4y}
\phi = \frac{A\sn(\beta x, m) \pm iB \cn(\beta x, m)\, \dn(\beta x, m)}
{1-|D|\sn^2(\beta x, m)}\,,
\ee
provided
\be\label{2.5y}
a = -(1+m)\beta ^2 < 0\,,~~b A^2 = -2(1-|D|)(|D|-m)\beta^2\,,~~
bB^2 = -2|D|\beta^2\,,~~m < |D| < 1\,.
\ee
It may be noted that such a solution is not valid in the m = 1 limit.
 
In the limit $m = 1$, we are led to the complex PT-invariant 
kink solution  
\be\label{2.9}
\phi = \frac{A\tanh(\beta x) \pm iB \sech^2(\beta x)}{1+D\tanh^2(\beta x)}\,,
\ee
provided
\be\label{2.10}
a = -2\beta ^2 < 0\,,~~b A^2 = 2(1+D)^2 \beta^2 > 0\,,~~bB^2 
= 2D\beta^2 > 0\,.
\ee

Is Eq. (\ref{2.9}) a new (hyperbolic) kink solution? Unfortunately, the answer 
to the question is no. It is apparent from Eq. (\ref{2.10})  that 
the solution (\ref{2.9}) is valid only if $D > 0$ and hence the arguments
given above about translation hold good. It also directly follows from a 
similar argument in the hyperbolic case. In particular,  
since $\phi^4$ model is a translation invariant system, hence if
$\hat{A}\tanh(\beta x)$ is a solution then so is 
$\hat{A}\tanh(\beta x+ia)$. It is then 
easily shown that the corresponding solution has precisely the form of the 
solution (\ref{2.9}) with
\be\label{2.12}
a = -(1+m) \beta^2\,,~~A = \sec^2(a) \hat{A}\,,~~B = \tan(a) \hat{A}\,,~~D = 
\tan^2(a) > 0\,,
\ee
with $a, \hat{A}$ satisfying Eq. (\ref{2.3}).
 
Summarizing, while we have obtained a new complex PT-invariant 
periodic kink solution of the $\phi^4$ field Eq. (\ref{I2}) as given by 
Eqs. (\ref{2.4y}) and (\ref{2.5y}), the corresponding
complex PT-invariant hyperbolic kink solution (\ref{2.9}) is merely a 
translation of the celebrated kink solution $\tanh(\beta x)$.    

It may be noted that even though we are calling it a complex kink solution, since 
the solution is only valid when both $a, b < 0$, effectively it is the solution
in the single well case. However, the way it has been derived and considering
its structure we are calling it a complex kink solution.  

\subsection{Connections Between The Complex Solutions of Several Real 
Nonlinear Equations}

Unlike the real solutions, there are novel connections between the various 
complex solutions of different models or even among the different solutions 
of the same model but for different values of the parameters. We now point 
out such connections between some of the well known models.

\begin{enumerate}

\item If for a given set of parameters $a, b$, $\phi_1$ is a complex solution
of the $\phi^4$ field Eq. (\ref{I2}), then $\pm i\phi_1$ is also a solution of
the same $\phi^4$ field Eq. (\ref{I2}) for the same value of parameter $a$ but
opposite value of the parameter $b$.

\item If for a given set of parameters $a, b, c$, $\phi_1$ is a complex 
solution of the field equation
\be\label{2.13}
\phi_{xx} = a \phi - b \phi^{2n+1} + c \phi^{4n+1}\,,~~n = 1, 2, 3,...
\ee
then for odd integer $n$, $\pm i\phi_1$ is also a solution of the same 
Eq. (\ref{2.13})
for the same values of the parameters $a$ and $c$ but opposite value of
the parameter $b$. On the other hand, for even integer $n$, 
$\pm i\phi_1$ is also a solution of the same Eq. (\ref{2.13}) for the
same values of the parameters $a, b, c$. Note that for $n = 1$, 
Eq. (\ref{2.13}) reduces to the celebrated $\phi^6$ field equation
\be\label{2.13h}
\phi_{xx} = a \phi - b \phi^{3} + c \phi^{5}\,.
\ee

\item If $u_1$ is a complex solution of the focusing mKdV equation \cite{dj,asegur} 
\be\label{2.14}
u_t(x,t) + u_{xxx} + 6 u^2 u_x = 0\,,
\ee
then $\pm i u_1$ must be the complex solution of the defocusing mKdV
equation
\be\label{2.15}
u_t(x,t) + u_{xxx} - 6 u^2 u_x = 0\,,
\ee
The converse is true as well.

\item If $\phi_1$ is a complex solution of the SG equation \cite{dj,asegur} 
\be\label{2.15a}
\phi_{xx} = \sin(\phi)\,,
\ee
then $\pm i\phi_1$ must be the complex solution of the SHG equation \cite{shg} 
\be\label{2.15b}
\phi_{xx} = \sinh(\phi)\,,
\ee
The converse is also true.

\item If for a given set of parameters $a, b$, $\phi_1$ is a complex 
solution of the DSG equation \cite{dsg} 
\be\label{2.16}
\phi_{xx} = (a/2) \sin(4\phi) - b \sin(2\phi)\,,
\ee
then for the {\it same} $a, b$, $\pm i\phi_1$ must be the complex 
solution of the DSHG equation \cite{razavi,khs,hks} 
\be\label{2.17}
\phi_{xx} = (a/2) \sinh(4\phi) - b \sinh(2\phi)\,,
\ee
The converse is true as well.

\end{enumerate}

As an illustration, we shall now obtain several new complex PT-invariant periodic 
solutions of the $\phi^4$ field Eq. (\ref{I2}). Further, in Sec. III and Sec. IV we shall 
use this connection both ways and obtain several complex
PT-invariant solutions of the attractive and repulsive mKdV Eqs. (\ref{2.14})
and (\ref{2.15}) as well as those of DSG and DSHG field Eqs. (\ref{2.16}) and 
(\ref{2.17}), respectively. 

\subsection{Superposed Complex PT-Invariant pulse solutions of $\phi^4$}
 
We now show that similar to the kink case, one also has novel complex 
PT-invariant periodic pulse solutions of the celebrated 
$\phi^4$ field Eq.~(\ref{I2}). In particular, we obtain 5 novel complex 
PT-invariant periodic pulse solutions of this equation,  
three with PT-eigenvalue +$1$ and two with PT-eigenvalue
$-1$. These are in addition to the two well known complex PT-invariant 
pulse solutions with PT-eigenvalue +$1$ of the $\phi^4$ field Eq. (\ref{I2}).  

{\bf Pulse Solution I} 

Let us first note that in view
of the $\phi_1 \leftrightarrow \pm i\phi_1$ connection mentioned above, 
since $\phi^4$ field 
Eq. (\ref{I2}) admits the complex PT-invariant periodic kink solution 
(\ref{2.4}), hence the same
$\phi^4$ field Eq. (\ref{I2}) must admit the complex PT-invariant periodic pulse
solution
\be\label{2.18}
\phi = \frac{B \cn(\beta x, m)\, \dn(\beta x, m) \pm i A \sn(\beta x, m)}
{1+D\sn^2(\beta x,m)}\,,
\ee
provided $a$ is same as in Eq. (\ref{2.5}) while $b$ has opposite value, i.e.
\be\label{2.19}
a = -(1+m)\beta ^2 < 0\,,~~b A^2 = -2(1+D)(m+D)\beta^2\,,~~bB^2 = -2D\beta^2\,.
\ee
Note that while $a < 0$, $b < 0$ or $b > 0$ depends on if $D > 0$ 
or if $-1 < D < -m$, respectively. Following the arguments of the previous
section, it is then clear that in case $D> 0$ and hence $b < 0$, the
solution (\ref{2.18}) is merely a translation of the pure imaginary solution
$i\hat{A} \sn(\beta x, m)$ while if $-1 < D < -m$ so that $b > 0$, then 
(\ref{2.18}) is a genuinely new solution which cannot be obtained by
translation from the pure imaginary solution $i\hat{A} \sn(\beta x, m)$.
Summarizing, we have discovered a new complex PT-invariant periodic pulse
solution with PT-eigenvalue +$1$ given by
\be\label{2.18x}
\phi = \frac{B \cn(\beta x, m)\, \dn(\beta x, m) \pm i A \sn(\beta x, m)}
{1-|D|\sn^2(\beta x,m)}\,,
\ee
provided 
\be\label{2.19x}
a = -(1+m)\beta ^2 < 0\,,~~b A^2 = 2(1-|D|)(|D|-m)\beta^2\,,
~~bB^2 = 2|D|\beta^2\,.
\ee

We would like to emphasize that this pulse solution (for $b > 0$) is an 
entirely new 
complex PT-invariant periodic pulse solution which is distinct from the 
two well known \cite{ks1} complex periodic pulse solutions 
of the $\phi^4$ field Eq. (\ref{I2}) which are only valid if $b < 0$. 
In particular, it is well known that \cite{ks1} 
\be\label{2.20}
\phi = A[\dn(\beta x, m) \pm i\sqrt{m} \sn(\beta x, m)]\,,
\ee
is an exact complex PT-invariant pulse solution of the $\phi^4$ field Eq. (\ref{I2}) 
provided
\be\label{2.21}
a = -(2m-1)/2 \beta^2\,,~~b A^2 = -\beta^2\,.
\ee
The second solution is \cite{ks1}
\be\label{2.22}
\phi = A\sqrt{m} [\cn(\beta x, m) \pm i \sn(\beta x, m)]\,,
\ee
provided
\be\label{2.23}
a = -(2-m)/2 \beta^2\,,~~b A^2 = -\beta^2\,.
\ee

Following the discussion for the kink case, it is clear that in the limit 
$m =1$, the solution (\ref{2.18}) is not an independent solution but 
merely a translation of the pure imaginary solution
$i\hat{A} \tanh(\beta x)$. 
 
We now show that apart from the complex periodic pulse solution (\ref{2.18x})
with PT-eigenvalue +$1$, there are four more novel complex PT-invariant 
periodic pulse solutions of the $\phi^4$ field Eq. (\ref{I2}), two with 
PT-eigenvalue +$1$ and two with PT-eigenvalue $-1$. 

{\bf Pulse Solutions II and III}

To motivate the solutions, we note that one of the well known periodic 
pulse solutions to $\phi^4$ field Eq. (\ref{I2}) is \cite{phi4} 
\be\label{2.27}
\phi = A \cn(\beta x,m)\,,
\ee
provided
\be\label{2.28}
b A^2 =  -2 m \beta^2\,,~~a= (2m-1) \beta^2\,.
\ee
Since $\phi^4$ model is a translation invariant system, hence if
$\cn(\beta x, m)$ is a solution then so is $\cn(\beta x \pm iK'(m)/2,m)$.
At this stage it is worth recalling the identity \cite{as} 
\be\label{2.30}
\cn[u \pm iK'(m)/2, m] = \frac{(1+\sqrt{m})^{1/2}[\cn(u, m) \mp i \sn(u, m)\,
\dn(u, m)}{m^{1/4} [1+\sqrt{m} \sn^2(u, m)]}\,.
\ee
Inspired by this identity, as in the kink case, we now inquire if there is 
a more general complex, PT-invariant periodic pulse 
solution of the $\phi^4$ field Eq. (\ref{I2}) with PT-eigenvalue +$1$. We start 
with the ansatz
\be\label{2.31}
\phi = \frac{A\cn(\beta x, m) \pm iB \sn(\beta x, m)\, \dn(\beta x, m)}
{1+D\sn^2(\beta x,m)}\,,
\ee
where $A, B, D, \beta$ have to be determined in terms of the parameters of the
model. After a lengthy algebra we find that indeed Eq. (\ref{2.31}) is a novel
complex PT-invariant periodic pulse solution of the $\phi^4$ field Eq. (\ref{I2}) 
provided
\be\label{2.32}
a = (2m-1)\beta ^2\,,~~b A^2 = -2(m+D)\beta^2\,,~~b B^2 = -2D(1+D)\beta^2\,.
\ee
From here we conclude that while the sign of $a$ depends on whether $m > (<) 1/2$, 
$b < 0$ if $D > 0$ while  $b > 0$ if $-1 < D < -m$.

It is reasonable to expect that at least
in the case $D> 0$ (and hence  $b < 0$), the solution (\ref{2.31}) may simply 
follow from
the real periodic pulse solution (\ref{2.27}) by translation. We now show that
this is indeed the case. 
As argued above, since $\phi^4$ model is a translation invariant system, hence if
$\hat{A} \cn(\beta x)$ is a solution of the $\phi^4$ field Eq. (\ref{I2})
then so is $\hat{A} \cn(\beta x +ia)$. On using the addition
formula for $\cn(x,m)$ given by \cite{as} 
\be\label{2.32b}
\cn(a+b, m) = \frac{\cn(a, m)\, \cn(b, m) - \sn(a, m)\, \dn(a, m)\, \sn(b, m)\, 
\dn(b, m)}{1-m\sn^2(a, m)\, \sn^2(b, m)}\,, 
\ee
and using Eq. (\ref{2.5d}) it immediately follows that in the case $D > 0$ and
hence $b < 0$, the corresponding solution has precisely the form of 
solution (\ref{2.31}) with
\be\label{2.32f}
A = \hat{A} \frac{1}{\cn(a, 1-m)}\,,~~B = -\hat{A}
\frac{\sn(a, 1-m) \dn(a, 1-m)}{\cn^2(a, 1-m)}\,,~~D 
= \frac{m \sn^2(a, 1-m)}{\cn^2(a, 1-m)} > 0\,,
\ee
and the condition (\ref{2.32}) takes the form 
\be\label{2.32g}
a = (2m-1) \beta^2\,,~~b \hat{A}^2 = -2m \beta^2\,,
\ee
which is precisely the condition under which the solution 
$\hat{A} \cn(\beta x, m)$ is satisfied. It must be emphasized 
that if instead one considers solution (\ref{2.31}) with $D < 0$, then 
such a solution cannot follow by translation from the solution (\ref{2.27}) 
since as is clear from Eq. (\ref{2.32f}), such a translation always 
gives $D > 0$. Thus we have shown that the $\phi^4$ field Eq. (\ref{I2}) 
admits yet another novel complex PT-invariant periodic pulse solution 
with PT-eigenvalue +$1$ as given by 
\be\label{2.31x}
\phi = \frac{A\cn(\beta x, m) \pm iB \sn(\beta x, m)\, \dn(\beta x, m)}
{1-|D|\sn^2(\beta x,m)}\,,
\ee
provided
\be\label{2.32x}
a = (2m-1)\beta ^2\,,~~b A^2 = 2(|D|-m)\beta^2\,,~~b B^2 
= 2|D|(1-|D|)\beta^2\,.
\ee

In view of the $\phi_1 \leftrightarrow \pm i\phi_1$ connection,  it then 
follows that the 
$\phi^4$ field Eq. (\ref{I2})  also admits the complex PT-invariant 
periodic pulse solution with PT-eigenvalue $-1$ given by
\be\label{2.31a}
\phi = \frac{B \sn(\beta x, m)\, \dn(\beta x, m) \pm i A\cn(\beta x, m)}
{1-|D|\sn^2(\beta x,m)}\,,
\ee
provided
\be\label{2.32a}
a = (2m-1)\beta ^2\,,~~b A^2 = -2(|D|-m)\beta^2\,,~~b B^2 
= -2|D|(1-|D|)\beta^2\,.
\ee
Observe that this is a novel pulse solution for which $a$ could be positive
(or negative) depending on whether $m< (>) 1/2$ while $b < 0$ since 
$-1 < D <-m$. 
We would like to emphasize that the pulse solution (\ref{2.31a}) 
is an entirely new 
complex PT-invariant periodic pulse solution with PT-eigenvalue $-1$, 
which is distinct from any of the  well
known \cite{ks1} complex periodic pulse solutions 
of the $\phi^4$ field Eq. (\ref{I2}). As far as we are aware of, all the known
pulse solutions of the $\phi^4$ field Eq. (\ref{I2}) have PT-eigenvalue +$1$
while this solution has PT-eigenvalue $-1$.

{\bf Fourth and Fifth Pulse Solutions}

To motivate the fourth novel complex periodic pulse solution, we note that 
the $\phi^4$ field Eq. (\ref{I2}) also admits another periodic pulse solution \cite{phi4} 
\be\label{2.36}
\phi = A \dn(\beta x,m)\,,
\ee
provided
\be\label{2.37}
b A^2 =  -2\beta^2\,,~~a = (2-m) \beta^2\,.
\ee
Now since $\phi^4$ model is a translation invariant system, hence if
$\hat{A} \dn(\beta x, m)$ is a solution then so is 
$\hat{A} \dn(\beta x \pm iK'(m)/2,m)$.
At this stage it is worth recalling the identity \cite{as} 
\be\label{2.38}
\dn[u \pm iK'(m)/2, m] = \frac{(1+\sqrt{m})^{1/2}[\dn(u, m) \mp i \sqrt{m} 
\sn(u, m)\, \cn(u, m)}{[1+\sqrt{m} \sn^2(u, m)]}\,.
\ee
Inspired by this identity, we now inquire if there is a more general 
complex, PT-invariant periodic pulse 
solution of the $\phi^4$ field Eq. (\ref{I2}). To that effect we start with the 
ansatz
\be\label{2.39}
\phi = \frac{A\dn(\beta x, m) \pm iB \sn(\beta x, m)\, \cn(\beta x, m)}
{1+D\sn^2(\beta x, m)}\,,
\ee
where $A, B, D, \beta$ have to be determined in terms of the parameters of the
model. After a lengthy algebra we find that indeed Eq. (\ref{2.39}) is a novel
complex PT-invariant periodic pulse solution of the $\phi^4$ field Eq. (\ref{I2}) 
with PT-eigenvalue +$1$ provided
\be\label{2.40}
a = (2-m)\beta ^2 > 0\,,~~b A^2 = -2(1+D)\beta^2\,,~~b B^2 = -2D(m+D)\beta^2\,.
\ee
From here we conclude that unlike the pulse solution (\ref{2.31}), this
pulse solution exists only if $a > 0$ and $b < 0$ irrespective of whether
$D > 0$ or if $-1 < D < -m$. 

It is then reasonable to expect that
in the case $D> 0$, the solution (\ref{2.39}) would follow from
the real periodic pulse solution (\ref{2.36}) by translation. We now show that
this is indeed the case.   
As argued above, since $\phi^4$ model is a translation invariant system, hence if
$\hat{A} \dn(\beta x)$ is a solution of the $\phi^4$ field Eq. (\ref{I2})
then so is $\hat{A} \dn(\beta x +ia)$. On using the addition
formula for $\dn(x,m)$ given by \cite{as} 
\be\label{2.39h}
\dn(a+b, m) = \frac{\dn(a, m)\, \dn(b, m) - m \sn(a, m)\, \cn(a, m)\, 
\sn(b, m)\, \cn(b, m)}{1-m\sn^2(a, m)\, \sn^2(b, m)}\,, 
\ee
and using Eq. (\ref{2.40}) it immediately follows that in the case $D > 0$, 
the corresponding solution has precisely the form of the solution 
(\ref{2.39}) with
\be\label{2.39f}
A = \hat{A} \frac{\dn(a, 1-m)}{\cn(a, 1-m)}\,,~~B = -m \hat{A}
\frac{\sn(a, 1-m)}{\cn^2(a, 1-m)}\,,~~
D = \frac{m \sn^2(a, 1-m)}{\cn^2(a, 1-m)} > 0\,,
\ee
and the condition (\ref{2.40}) takes the form 
\be\label{2.39g}
a = (2-m) \beta^2\,,~~b \hat{A}^2 = -2 \beta^2\,,
\ee
which is precisely the condition under which the solution 
$\hat{A} \dn(\beta x, m)$ is satisfied. It must be emphasized once again
that if instead one considers the solution (\ref{2.39}) with $-1 < D < -m$, 
then such a solution cannot follow by translation  since, as
is clear from Eq. (\ref{2.39f}), such a translation always gives $D > 0$.   
Thus for $ -1 < D < -m$, we have yet another novel complex PT-invariant 
periodic pulse solution of the $\phi^4$ field Eq. (\ref{I2}) with PT-eigenvalue 
+$1$ and is given by
\be\label{2.39x}
\phi = \frac{A\dn(\beta x, m) \pm iB \sn(\beta x, m)\, \cn(\beta x, m)}
{1-|D|\sn^2(\beta x, m)}\,,
\ee
provided
\be\label{2.40x}
a = (2-m)\beta ^2 > 0\,,~~b A^2 = -2(1-|D|)\beta^2\,,
~~b B^2 = -2|D|(|D|-m)\beta^2\,.
\ee

In view of the $\phi_1 \leftrightarrow \pm i\phi_1$ symmetry, it then follows 
that the $\phi^4$ field Eq. (\ref{I2}) also admits a complex PT-invariant 
periodic pulse solution with PT-eigenvalue $-1$ given by
\be\label{2.39a}
\phi = \frac{B \sn(\beta x, m)\, \cn(\beta x, m) \pm i A\dn(\beta x, m)}
{1-|D|\sn^2(\beta x,m)}\,,
\ee
provided
\be\label{2.40a}
a = (2-m)\beta ^2 > 0\,,~~b A^2 = 2(1-|D|)\beta^2\,,~~b B^2 
= 2|D|(|D|-m)\beta^2\,.
\ee
Thus, we now have discovered two new periodic pulse solutions of the 
$\phi^4$ field Eq. (\ref{I2}) with PT-eigenvalue $-1$ as given by Eqs. (\ref{2.31a})
and (\ref{2.39a}).

In the limit $m = 1$, both the complex PT-invariant periodic pulse solutions
(\ref{2.31}) and (\ref{2.39}) with PT-eigenvalue +$1$ go over to the complex 
PT-invariant (hyperbolic) pulse solution  
\be\label{2.44}
\phi = \frac{A\sech(\beta x) \pm iB \sech(\beta x)\, \tanh(\beta x)}
{1+D\tanh^2(\beta x)}\,,
\ee
provided
\be\label{2.45}
a = \beta ^2 > 0\,,~~b A^2 = -2(1+D)^2 \beta^2\,,~~bB^2 = 
-2D(1+D)\beta^2\,.
\ee

Is Eq. (\ref{2.44}) a new (hyperbolic) pulse solution? Unfortunately, the answer 
to the question is no. It is in fact clear from Eq. (\ref{2.45})  that 
the solution (\ref{2.44}) is valid only if $D > 0$ (note $D > -1$ so as
to avoid any singularity) and hence the arguments
given above about translation hold good. It also directly follows from a 
similar argument in the hyperbolic case. In particular,  
since $\phi^4$ model is a translation invariant system, hence if
$\hat{A}\sech(\beta x)$ is a solution then so is 
$\hat{A}\sech(\beta x+ia)$. It is then 
easily shown that the corresponding solution has precisely the form of the 
solution (\ref{2.44}) with
\be\label{2.45b}
a = \beta^2\,,~~A = \sec^2(a) \hat{A}\,,~~B = \tan(a) \sec(a) \hat{A}\,,~~D = 
\tan^2(a) > 0\,,
\ee
with $a, \hat{A}$ satisfying Eq. (\ref{2.3}).
 
Summarizing, apart from the two well known complex PT-invariant periodic pulse 
solutions (\ref{2.20}) and (\ref{2.22}) with PT-eigenvalue +$1$, we have 
discovered
five new novel complex periodic PT-invariant pulse
solutions, three with PT-eigenvalue +$1$ while two are with PT-eigenvalue $-1$. 
Note, however, that the corresponding hyperbolic solutions are merely
translations of the well known real kink and pulse solutions.

\section{Solutions of Both Focusing and Defocusing mKdV}

Let us consider the defocusing mKdV equation \cite{dj,asegur}
\be\label{4.1}
u_t + u_{xxx} - 6 u^2 u_x = 0\,.
\ee
Note that if $u_1$ is a complex solution of the defocusing mKdV 
Eq. (\ref{4.1}), then for the {\it same} values of the parameters 
$\pm i u_1$ is a solution of the focusing mKdV equation
\be\label{4.2}
u_t + u_{xxx} + 6 u^2 u_x = 0\,.
\ee
Note that this connection applies both ways. Using the
various novel complex solutions obtained in the last section, 
we can then immediately write down new  solutions of both
the focusing and the defocusing mKdV equations.

{\bf Complex PT-invariant Periodic Solutions of Defocusing mKdV}

Let us start from the defocusing mKdV Eq. (\ref{4.1}). On using 
$y = x-vt$, and integrating once, Eq. (\ref{4.1}) takes the form
\be\label{4.3}
u_{yy} = v u + 2 u^3 + k\,,
\ee
where $k$ is an integration constant. Some of the well known periodic 
solutions to the defocusing mKdV Eq. (\ref{4.3}) are \cite{dj,asegur}
$u = A\sn(\beta y, m)$ and the complex PT-invariant periodic solutions
$A[\sn(\beta x, m) \pm i \cn(\beta x, m)]$ as well as
 $A[\sn(\beta x, m) \pm i \dn(\beta x, m)]$ with PT-eigenvalue $-1$ 
\cite{ks1}. Using the results of the last
section, we now show that it has three additional complex PT-invariant 
periodic solutions, two with PT-eigenvalue +$1$ and one with PT-eigenvalue
$-1$. 

{\bf Solution I:}
Following the results of the last section, it is easily shown that 
\be\label{4.4}
u(y=x-vt) = \frac{B \cn(\beta y, m) \dn(\beta y, m) \pm i \sn(\beta y, m)}
{1-|D|\sn^2(\beta y, m)}\,,
\ee
is an exact complex PT-invariant periodic solution with PT-eigenvalue +$1$ 
of Eq. (\ref{4.3}) with
\be\label{4.5}
k = 0\,,~~v = -(1+m)\beta^2 < 0\,,~~ A^2 = (1-|D|)(|D|-m)\beta^2\,,
~~B^2 = |D|\beta^2\,.
\ee

{\bf Solution II:} 
The other complex PT-invariant periodic solution with
PT-eigenvalue +$1$ of the defocusing mKdV Eq. (\ref{4.3}) is
\be\label{4.6}
u = \frac{A \cn(\beta y, m) \pm iB \sn(\beta y, m) \dn(\beta y, m)}
{1-|D|\sn^2(\beta y, m)}\,,
\ee
provided
\be\label{4.7}
k = 0\,,~~v = (2m-1)\beta^2\,,~~ A^2 = (|D|-m)\beta^2\,,
~~ B^2 = |D|(1-|D|)\beta^2\,.
\ee

{\bf Solution III:}
The third solution is the complex PT-invariant periodic solution with
PT-eigenvalue $-1$
\be\label{4.8}
u = \frac{B \sn(\beta y, m) \cn(\beta y, m) \pm i A\dn(\beta y, m)}
{1-|D|\sn^2(\beta y, m)}\,,
\ee
provided
\be\label{4.9}
k = 0\,,~~v = (2-m)\beta^2 > 0\,,~~ A^2 = (1-|D|)\beta^2\,,~~ 
B^2 = |D|(|D|-m)\beta^2\,.
\ee

{\bf Complex PT-invariant Periodic Solutions of Focusing mKdV}

Let us start from the focusing mKdV Eq. (\ref{4.2}). On using 
$y = x-vt$, and integrating once, Eq. (\ref{4.2}) takes the form
\be\label{4.10}
u_{yy} = v u - 2 u^3 + k\,,
\ee
where $k$ is an integration constant.  In addition to its
well known periodic real solutions $A \cn(\beta x, m)$ and 
$A\dn(\beta x, m)$ \cite{dj,asegur}, 
it also has two complex periodic solutions with PT-eigenvalue +$1$ given
by $\cn(\beta x, m) \pm i \sn(\beta x, m)$ and $\dn(\beta x, m) \pm i
\sn(\beta x, m)$ \cite{ks1}. On using the 
$u_1 \leftrightarrow \pm i u_1$ 
connection, one can immediately write down its solutions from the known 
solutions 
of the focusing mKdV. In particular it has three novel 
complex PT-invariant periodic solutions,
two with PT-eigenvalue $-1$ and one with PT-eigenvalue +$1$.

{\bf Solution I:}
Following the solutions of the defocusing  mKdV case, it follows that 
\be\label{4.11}
u(y=x-vt) = \frac{A \sn(\beta y, m) \pm iB \cn(\beta y, m) \dn(\beta y, m)} 
{1-|D|\sn^2(\beta y, m)}\,,
\ee
is an exact complex PT-invariant periodic solution
of the focusing mKdV Eq.~(\ref{4.10}) with PT-eigenvalue $-1$
\be\label{4.12}
k = 0\,,~~v = -(1+m)\beta^2 < 0\,,~~ A^2 = (1-|D|)(|D|-m)\beta^2\,,
~~B^2 = |D|\beta^2\,.
\ee

{\bf Solution II:}
The other complex PT-invariant periodic solution with
PT-eigenvalue +$1$ of the focusing mKdV Eq. (\ref{4.10}) is
\be\label{4.13}
u = \frac{B \sn(\beta y, m) \dn(\beta y, m) \pm iA \cn(\beta y, m)}
{1-|D|\sn^2(\beta y, m)}\,,
\ee
provided
\be\label{4.14}
k = 0\,,~~v = (2m-1)\beta^2\,,~~ A^2 = (|D|-m)\beta^2\,,
~~ B^2 = |D|(1-|D|)\beta^2\,.
\ee

{\bf Solution III:}
The third solution with PT-eigenvalue +$1$ is
\be\label{4.8a}
u = \frac{A \dn(\beta y, m) \pm iB \sn(\beta y, m) \cn(\beta y, m)}
{1-|D|\sn^2(\beta y, m)}\,,
\ee
provided
\be\label{4.9a}
k = 0\,,~~v = (2-m)\beta^2 > 0\,,~~ A^2 = (1-|D|)\beta^2\,,~~ 
B^2 = |D|(|D|-m)\beta^2\,.
\ee

\section{The Complex Solutions of DSHG and DSG}

In this section we exploit the $\phi_1 \leftrightarrow \pm i\phi_1$ 
connection discussed in 
Section 2 which relates the solutions of the DSHG and the DSG equations (see 
Eqs. (\ref{2.16}) and (\ref{2.17}) and discussion therein) and obtain several 
new complex PT-invariant solutions of these equations.

Let us first start with the DSHG Eq. (\ref{2.16}). On using the substitution
$u = \tanh(\phi)$ it is easily shown that the DSHG Eq. (\ref{2.16}) takes
the form
\be\label{3.1}
(1-u^2)u_{xx} +2u u_{x}^{2} = 2(a-b) u +2(a+b) u^3\,.
\ee
As has been recently shown by us \cite{ks2}, this equation admits the 
complex PT-invariant periodic kink solution with (PT-eigenvalue $-1$)
\be\label{3.2}
u = A[\sqrt{m} \sn(\beta x, m) \pm i \dn(\beta x, m)]\,,
\ee
provided
\bea\label{3.3}
&&\beta^2 = \frac{4(b^2-a^2)}{[(2m-1)a +\sqrt{a^2-4m(1-m)b^2}]}\,,
\nonumber \\
&&A^2 = \frac{[(2m-1)b - \sqrt{a^2-4m(1-m)b^2}]}{(a+b)}\,.
\eea

In view of the $u_1 \leftrightarrow \pm i u_1$ symmetry, we then immediately 
obtain a new PT-invariant pulse solution of the DSG Eq. (\ref{2.17}). To that
purpose, let us first note that if we make the substitution $v = \tan(\phi)$
then the DSG Eq. (\ref{2.17}) goes over to  
\be\label{3.4}
(1+v^2)v_{xx} -2v v_{x}^{2} = 2(a-b) v - 2(a+b) v^3\,.
\ee
Observe that under the transformation $u \rightarrow \pm iv$, the DSHG 
Eq. (\ref{3.2}) goes over to the DSG Eq. (\ref{3.4}). Thus for the {\it same}
set of parameters as given by Eq. (\ref{3.3}) we  immediately
obtain the new complex PT-invariant pulse solution of the DSG 
Eq. (\ref{3.4}) with PT-eigenvalue +$1$  
\be\label{3.5}
v = A[\dn(\beta x, m) \pm i \sqrt{m} \sn(\beta x, m)]\,. 
\ee

Similarly, corresponding to the other complex periodic PT-invariant kink 
solution of DSHG Eq. (\ref{2.16}) (and hence (\ref{3.1})) obtained in \cite{ks2}, 
we find the following complex PT-invariant periodic pulse solution of DSG with 
PT-eigenvalue +$1$
\be\label{3.6}
v = A\sqrt{m} [\cn(\beta x, m) \pm i \sn(\beta x, m)]\,,
\ee
provided  
\bea\label{3.7}
&&\beta^2 = \frac{4(b^2-a^2)}{[(2-m)a +(2-m)(1-m)b 
+m\sqrt{m^2 a^2+4(1-m)b^2}]}\,,
\nonumber \\
&&A^2 = \frac{[(2-m)b - \sqrt{m^2 a^2+4(1-m)b^2}]}{(a+b)}\,.
\eea

It may be noted that all the above complex PT-invariant kink solutions of DSHG 
were with PT-eigenvalue $-1$ and (hence the corresponding pulse solutions 
of DSG with PT-eigenvalue +$1$) are only valid if $b > a > 0$.

So far we have presented several complex PT-invariant solutions of the DSHG 
with PT-eigenvalue $-1$ and hence of the DSG with PT-eigenvalue +$1$ and 
it turned out that all these solutions are valid if $b > a > 0$. 
We now present a few complex PT-invariant pulse solutions of DSHG 
Eq. (\ref{3.1}) and hence Eq. (\ref{2.16}) with PT-eigenvalue +$1$ 
(and thus complex PT-invariant solutions of DSG with PT-eigenvalue $-1$).
We will see that all of these solutions are only valid in the case $a < b < 0$.

As has been recently shown by us \cite{ks2}, the DSHG Eq. (\ref{3.1})
 admits the 
complex PT-invariant periodic pulse solution (with PT-eigenvalue +$1$)
\be\label{3.12}
u = A[\dn(\beta x, m) \pm i\sqrt{m} \sn(\beta x, m)]\,,
\ee
provided
\bea\label{3.13}
&&\beta^2 = \frac{4(a+b)[(2m-1)b -\sqrt{a^2-4m(1-m)b^2}]}{a +4m(1-m) b
+(2m-1)\sqrt{a^2 -4m(1-m) b^2}}\,,
\nonumber \\
&&A^2 = \frac{[-(2m-1)b + \sqrt{a^2-4m(1-m)b^2}]}{(a+b)}\,.
\eea
This solution immediately implies the existence of the complex PT-invariant
periodic kink solution of the DSG Eq. (\ref{3.4})
\be\label{3.14}
v = A[\sqrt{m} \sn(\beta x, m) \pm i\dn(\beta x, m)]\,,
\ee
provided Eq. (\ref{3.13}) is satisfied.

The DSHG Eq. (\ref{3.1}) also admits \cite{ks2} another complex PT-invariant
periodic pulse solution
\be\label{3.15}
u = A\sqrt{m} [\cn(\beta x, m) \pm i \sn(\beta x, m)]\,,
\ee
provided
\bea\label{3.16}
&&\beta^2 = \frac{4(a+b)[(2-m)b -\sqrt{m^2 a^2 +4(1-m)b^2}]}{m^2 a 
-4(1-m) b +(2-m)\sqrt{m^2 a^2 +4(1-m) b^2}}\,,
\nonumber \\
&&A^2 = \frac{[-(2-m)b + \sqrt{m^2 a^2 +4(1-m)b^2}]}{m(a+b)}\,.
\eea
This solution immediately implies the existence of the complex PT-invariant
periodic kink solution of the DSG Eq. (\ref{3.4})
\be\label{3.17}
v = A\sqrt{m} [\sn(\beta x, m) \pm i\cn(\beta x, m)]\,,
\ee
provided Eq. (\ref{3.16}) is satisfied.

\section{Complex PT-invariant Solutions of a Coupled $\phi^4$ Model}

We now consider a coupled $\phi^4$ model which has been discussed before
\cite{ks2,ks3} and show that it admits a large number of novel
complex PT-invariant periodic solutions.
The coupled model has the static equations of motion
\be\label{5.1}
\phi_{xx} = a_1 \phi + b_1 \phi^3 + \alpha \phi \psi^2\,,
\ee  
\be\label{5.2}
\psi_{xx} = a_2 \psi + b_2 \psi^3 + \alpha \psi \phi^2\,. 
\ee  
In Sec. II we have obtained six novel complex PT-invariant periodic solutions
of the (uncoupled) $\phi^4$ field Eq. (\ref{I2}). In this section we will obtain
solutions of the coupled model as given by Eqs. (\ref{5.1}) and (\ref{5.2})
using these six solutions. It is then clear that there will be in general
twenty one distinct solutions of this coupled model. Out of these twenty one,
six solutions will be such that $\phi$ and $\psi$ will have similar form while
in fifteen solutions, $\phi$ and $\psi$ will have different form. 
Let us discuss these solutions one by one.

{\bf Solution I:}
One can show that 
\be\label{5.3}
\phi = \frac{A\sn(\beta x, m) \pm iB \cn(\beta x, m) \dn(\beta x, m)}{1-|D|
\sn^2(\beta x, m)}\,,~~  
\psi = \frac{E\sn(\beta x, m) \pm iF \cn(\beta x, m) \dn(\beta x, m)}{1-|D|
\sn^2(\beta x, m)}\,,   
\ee
is an exact solution of the coupled Eqs. (\ref{5.1}) and (\ref{5.2}) 
provided
\be\label{5.4}
a_1 = a_2 = -(1+m) \beta^2\,,~~A/B = E/F = \sqrt{(1-|D|)(|D|-m)}{|D|}\,,~~
b_1 A^2 + \alpha E^2 = b_2 E^2 +\alpha A^2 = -2(1-|D|)(|D|-m)\beta^2\,.
\ee
Note that $-1 < D < -m$. On solving we find that 
\be\label{5.5}
A^2 = \frac{2(1-|D|)(|D|-m)(\alpha -b_2)}{b_1 b_2 - \alpha^2}\,,~~
E^2 = \frac{2(1-|D|)(|D|-m)(\alpha -b_1)}{b_1 b_2 - \alpha^2}\,,
\ee
unless $b_1 = b_2 = \alpha < 0$. In that case we cannot determine $A, E$ and
hence $B, F$ separately but they must satisfy the constraint
\be\label{5.6}
b_1(A^2+E^2) = -2(1-|D|)(|D|-m)\beta^2\,,~~b_1(B^2+F^2) = -2|D| \beta^2\,.
\ee

{\bf Solution II:}
The solution of the form 
\be\label{5.7}
\phi = \frac{B \cn(\beta x, m) \dn(\beta x, m) \pm iA \sn(\beta x, m)}{1-|D|
\sn^2(\beta x, m)}\,,~~  
\psi = \frac{F \cn(\beta x, m) \dn(\beta x, m) \pm iE \sn(\beta x, m)}{1-|D|
\sn^2(\beta x, m)}\,,
\ee
can now be immediately obtained by comparing it with Solution I. In 
particular, notice that Solution II can be obtained from Solution I by 
changing $A$ and $E$ to $\pm iA$ and  $\pm iE$, respectively, while changing 
$B$ and $F$ to $\mp iB$ and $\mp iF$, respectively. In particular, one can 
show that (\ref{5.7}) is an exact solution of the coupled Eqs. (\ref{5.1}) 
and (\ref{5.2}) provided
\be\label{5.8}
a_1 = a_2 = -(1+m) \beta^2\,,~~A/B = E/F  = \sqrt{\frac{(1-|D|)(|D|-m)}{|D|}}\,,~~
b_1 A^2 + \alpha E^2 = b_2 E^2 +\alpha A^2 = 2(1-|D|)(|D|-m)\beta^2\,.
\ee
On solving we find that 
\be\label{5.9}
A^2 = \frac{2(1-|D|)(|D|-m)(b_2-\alpha)}{b_1 b_2 - \alpha^2}\,,~~
E^2 = \frac{2(1-|D|)(|D|-m)(b_1-\alpha)}{b_1 b_2 - \alpha^2}\,,
\ee
unless $b_1 = b_2 = \alpha > 0$. In that case we cannot determine $A, E$ and
hence $B, F$ separately but they must satisfy the constraint
\be\label{5.9b}
b_1(A^2+E^2) = 2(1-|D|)(|D|-m)\beta^2\,,~~b_1(B^2+F^2) = 2|D| \beta^2\,.
\ee

{\bf Solution III:}
The solution of the form  
\be\label{5.10}
\phi = \frac{A \sn(\beta x, m) \pm iB \cn(\beta x, m) \dn(\beta x, m)}{1-|D|
\sn^2(\beta x, m)}\,,~~  
\psi = \frac{F \cn(\beta x, m) \dn(\beta x, m) \pm iE \sn(\beta x, m)}{1-|D|
\sn^2(\beta x, m)}\,,
\ee
can also be immediately obtained by comparing it with Solution I. In 
particular, Solution III is obtained from Solution I by changing $E$
to $\pm iE$  and changing $F$ to $\mp iF$ (while $A, B$
remain unaltered). In particular, one can show that (\ref{5.10})
 is an exact solution of the coupled Eqs. (\ref{5.1}) and (\ref{5.2}) 
provided
\bea\label{5.11}
a_1 = a_2 = -(1+m) \beta^2\,,~~A/B = -E/F  
= \sqrt{\frac{(1-|D|)(|D|-m)}{|D|}}\, \nonumber \\ 
b_1 A^2 - \alpha E^2 = \alpha A^2 - b_2 E^2 
 = -2(1-|D|)(|D|-m)\beta^2\,, ~~~ 
b_1 B^2 - \alpha F^2 = \alpha B^2 - b_2 F^2 = -2|D| \beta^2\,.
\eea
On solving we find that 
\bea\label{5.12}
&&A^2 = \frac{2(1-|D|)(|D-m)(\alpha -b_2)}{b_1 b_2 - \alpha^2}\,,~~
E^2 = \frac{2(1-|D|)(|D|-m)(b_1-\alpha)}{b_1 b_2 - \alpha^2}\,,
\nonumber \\
&&B^2 = \frac{2 |D| (\alpha-b_2)}{b_1 b_2 - \alpha^2}\,,~~
F^2 = \frac{2 |D| (b_1-\alpha)}{b_1 b_2 - \alpha^2}\,,
\eea
unless $b_1 = b_2 = \alpha$. In that case we cannot determine $A, E$ and
hence $B, F$ separately but they must satisfy the constraint
\be\label{5.13}
b_1(A^2-E^2) = -2 (1-|D|)(|D|-m)\beta^2\,,~~b_1(B^2-F^2) = -2 |D| \beta^2\,.
\ee

{\bf Solution IV:}
One can show that 
\be\label{5.14}
\phi = \frac{A\cn(\beta x, m) \pm iB \sn(\beta x, m) \dn(\beta x, m)}{1-|D|
\sn^2(\beta x, m)}\,,~~  
\psi = \frac{E\cn(\beta x, m) \pm iF \sn(\beta x, m) \dn(\beta x, m)}{1-|D|
\sn^2(\beta x, m)}\,,  
\ee
is an exact solution of the coupled Eqs. (\ref{5.1}) and (\ref{5.2}) 
provided
\be\label{5.15}
a_1 = a_2 = (2m-1) \beta^2\,,~~A/B = E/F = \sqrt{(|D|-m)}{(1-|D|) |D|}\,,~~
b_1 A^2 + \alpha E^2 = b_2 E^2 +\alpha A^2 = 2 (|D|-m)\beta^2\,.
\ee
On solving we find that 
\be\label{5.16}
A^2 = \frac{2(|D|-m)(b_2-\alpha)}{b_1 b_2 - \alpha^2}\,,~~
E^2 = \frac{2(|D|-m)(b_1-\alpha)}{b_1 b_2 - \alpha^2}\,,
\ee
unless $b_1 = b_2 = \alpha > 0$. In that case we cannot determine $A, E$ and
hence $B, F$ seperately but they must satisfy the constraint
\be\label{5.17d}
b_1(A^2+E^2) = 2(|D|+m)\beta^2\,,~~b_1(B^2+F^2) = -2|D|(1-|D|) \beta^2\,.
\ee

Two more solutions can be immediately obtained from Solution IV
by exactly following the arguments used in obtaining Solutions II 
and III from Solution I. In particular, in one solution,  
$A$ and $E$ go over to
$\pm A$ and $\pm E$, respectively, while $B$ and $F$ go over to 
$\mp iB$ and $\mp iF$, respectively. In another solution we only let $E$
and $F$ go over to $\pm iE$ and $\mp iF$, respectively, while keeping A and
B unchanged. In each case Eq. (\ref{5.15}) and hence Eqs. (\ref{5.16}) and 
(\ref{5.17}) get suitably modified. We therefore do not write them down
here explicitly.

{\bf Solution V:}
Yet another solution of the coupled Eqs. (\ref{5.1}) and (\ref{5.2}) is 
\be\label{5.18}
\phi = \frac{A\dn(\beta x, m) \pm iB \sn(\beta x, m) \cn(\beta x, m)}{1-|D|
\sn^2(\beta x, m)}\,,~~  
\psi = \frac{E\dn(\beta x, m) \pm iF \sn(\beta x, m) \cn(\beta x, m)}{1-|D|
\sn^2(\beta x, m)}\,,  
\ee
provided
\be\label{5.19}
a_1 = a_2 = (2-m) \beta^2\,,~~A/B = E/F = \sqrt{(1-|D|)}{(|D|-m) |D|}\,,~~
b_1 A^2 + \alpha E^2 = b_2 E^2 +\alpha A^2 = -2 (1-|D|)\beta^2\,.
\ee
On solving we find that 
\be\label{5.20}
A^2 = -\frac{2(1-|D|)(b_2-\alpha)}{b_1 b_2 - \alpha^2}\,,~~
E^2 = -\frac{2(1-|D|)(b_1-\alpha)}{b_1 b_2 - \alpha^2}\,,
\ee
unless $b_1 = b_2 = \alpha < 0$. In that case we cannot determine $A, E$ and
hence $B, F$ seperately but they must satisfy the constraint
\be\label{5.21}
b_1(A^2+E^2) = -2(1-|D|)\beta^2\,,~~b_1(B^2+F^2) = -2|D|(|D|-m) \beta^2\,.
\ee

Two more solutions can be immediately obtained from Solution V
by exactly following the arguments used in obtaining Solutions II and III
from Solution I. In particular, in one solution, $A$ and $E$ go over to
$\pm A$ and $\pm E$, respectively, while $B$ and $F$ go over to 
$\mp iB$ and $\mp iF$, respectively. In another solution we only let $E$
and $F$ go over to $\pm iE$ and $\mp iF$, respectively, while keeping A and
B unchanged. In each case Eq. (\ref{5.15}) and hence Eqs. (\ref{5.16}) and 
(\ref{5.17}) get suitably modified. We therefore do not write them out here 
explicitly.

{\bf Solution VI:}
Yet another solution of the coupled Eqs. (\ref{5.1}) and (\ref{5.2}) is 
\be\label{5.22}
\phi = \frac{A\sn(\beta x, m) \pm iB \cn(\beta x, m) \dn(\beta x, m)}{1-|D|
\sn^2(\beta x, m)}\,,~~  
\psi = \frac{E\cn(\beta x, m) \pm iF \sn(\beta x, m) \dn(\beta x, m)}{1-|D|
\sn^2(\beta x, m)}\,,  
\ee
provided
\be\label{5.23}
\frac{A^2}{B^2} = \frac{(1-|D|)(|D|-m)}{|D|}\,,~~ \frac{AE}{BF} = 
-\frac{(|D|-m)}{|D|}\,,
\ee
\be\label{5.24}
m \alpha B^2 = |D|[a_2 -(2m-1)\beta^2]\,,~~m b_1 B^2 = |D|[a_1+(1-m)\beta^2]\,,
\ee
\be\label{5.25}
m \alpha E^2 = (|D|-m)[a_1 +(1+m)\beta^2]\,,~~m b_2 E^2 
= (|D|-m) [a_2 +\beta^2]\,.
\ee
On solving these equations we find that 
\be\label{5.26}
\beta^2 = \frac{\alpha a_2 - b_2 a_1}{(1+m) b_2 - \alpha} =
\frac{b_1 a_2 - \alpha a_1}{(2m-1) b_1 +(1-m) \alpha}\,.
\ee
This gives a constraint between the parameters $a_1, a_2, b_1, b_2, \alpha$
\be\label{5.27}
[a_1 -(1-m)a_2]\alpha^2 -2m (a_1 b_2 + a_2 b_1)\alpha +b_1 b_2 [(2m-1)a_1
+(1+m)a_2] = 0\,.
\ee

By following similar logic as in obtaining Solutions II and III from Solution I, we
can immediately obtain three more solutions from Solution VI. In one solution, one
changes $A$ and $E$ to $\pm iA$ and $\pm iE$, respectively, while changing
$B$ and $F$ to $\mp iB$ and $\mp iF$, respectively. In another solution
one only changes $A$ and $B$ to $\pm iA$ and $\mp iB$, respectively, while
keeping $E$ and $F$ unchanged.  In yet another solution one only changes $E$ 
and $F$ to $\pm iE$ and $\mp iF$, respectively, while keeping $A$ and $B$ unchanged.

{\bf Solution VII:}
Yet another solution of the coupled Eqs. (\ref{5.1}) and (\ref{5.2}) is 
\be\label{5.28}
\phi = \frac{A\sn(\beta x, m) \pm iB \cn(\beta x, m) \dn(\beta x, m)}{1-|D|
\sn^2(\beta x, m)}\,,~~  
\psi = \frac{E\dn(\beta x, m) \pm iF \sn(\beta x, m) \cn(\beta x, m)}{1-|D|
\sn^2(\beta x, m)}\,,  
\ee
provided
\be\label{5.29}
\frac{A^2}{B^2} = \frac{(1-|D|)(|D|-m)}{|D|}\,,~~ \frac{AE}{BF}  
= \frac{(1-|D|+1)}{|D|}\,,
\ee
\be\label{5.30}
 \alpha B^2 = -D[a_2 -(2-m)\beta^2]\,,~~ b_1 B^2 = -D[a_1-(1-m)\beta^2]\,,
\ee
\be\label{5.31}
\alpha E^2 = -(1-|D|)[a_1 +(1+m)\beta^2]\,,~~ b_2 E^2 
= -(1-|D|) [a_2 +m \beta^2]\,.
\ee
On solving these equations we find that 
\be\label{5.32}
\beta^2 = \frac{\alpha a_2 - b_2 a_1}{(1+m) b_2 - m\alpha} =
\frac{b_1 a_2 - \alpha a_1}{(2-m) b_1 -(1-m) \alpha}\,.
\ee
This gives a constraint between the parameters $a_1, a_2, b_1, b_2, \alpha$
\be\label{5.33}
[m a_1 +(1-m)a_2]\alpha^2 -2 (a_1 b_2 + a_2 b_1)\alpha +b_1 b_2 [(2-m)a_1
+(1+m)a_2] = 0\,.
\ee

By following similar logic as in obtaining Solutions II and III from Solution I, we 
can again obtain three more solutions from Solution VII. In one solution, one
changes $A$ and $E$ to $\pm iA$ and $\pm iE$, respectively, while changing
$B$ and $F$ to $\mp iB$ and $\mp iF$, respectively. In another solution one only 
changes $A$ and $B$ to $\pm iA$ and $\mp iB$, respectively, while keeping $E$ 
and $F$ unchanged. In yet another solution one only changes $E$ and $F$ to 
$\pm iE$ and $\mp iF$, respectively, while keeping $A$ and $B$ unchanged.  

{\bf Solution VIII:}
Yet another solution of the coupled Eqs. (\ref{5.1}) and (\ref{5.2}) is 
\be\label{5.34}
\phi = \frac{A\cn(\beta x, m) \pm iB \sn(\beta x, m) \dn(\beta x, m)}{1-|D|
\sn^2(\beta x, m)}\,,~~  
\psi = \frac{E\dn(\beta x, m) \pm iF \sn(\beta x, m) \cn(\beta x, m)}{1-|D|
\sn^2(\beta x, m)}\,,  
\ee
provided
\be\label{5.35}
\frac{A^2}{B^2} = \frac{(|D|-m)}{|D|(1-|D|)}\,,~~ \frac{AE}{BF}  
= -\frac{1}{|D|}\,,
\ee
\be\label{5.36}
b_1 A^2 + \alpha E^2 = -a_1 +(2|D|-1)\beta^2\,,~~b_2 E^2 +\alpha A^2 
= -a_2 +(2|D|-m)\beta^2\,.
\ee
On solving we find that 
\be\label{5.37}
A^2 = \frac{(\alpha a_2 - b_2 a_1)}{(b_1 b_2 - \alpha^2)} 
-\frac{(2|D|-m)\alpha -(2|D|-1)b_2}{b_1 b_2 - \alpha^2}\,,
\ee
\be\label{5.38}
E^2 = \frac{(\alpha a_1 - b_1 a_2)}{(b_1 b_2 - \alpha^2)} 
-\frac{(2|D|-1)\alpha -(2|D|-m)b_1}{b_1 b_2 - \alpha^2}\,,
\ee
unless $b_1 = b_2 = \alpha$. In that case we cannot determine $A, E$ and
hence $B, F$ seperately but they must satisfy the constraint
\be\label{5.17}
b_1(A^2 - \frac{|D|-m}{1-|D|} E^2) =  2(|D|-m)\beta^2\,,~~
b_1(B^2 -\frac{1-|D|}{|D|-m} F^2) = 2|D|(1-|D|) \beta^2\,.
\ee

By following similar logic as used in obtaining Solutions II and III from Solution I, 
we can again obtain three more solutions from Solution VII. In one solution, one
changes $A$ and $E$ to $\pm iA$ and $\pm iE$, respectively, while changing
$B$ and $F$ to $\mp iB$ and $\mp iF$, respectively. In another solution one only 
changes $A$ and $B$ to $\pm iA$ and $\mp iB$, respectively, while keeping $E$ 
and $F$ unchanged.  In yet another solution one only changes $E$ and $F$ to 
$\pm iE$ and $\mp iF$, respectively, while keeping $A$ and $B$ unchanged.  

\section{Summary and Open Problems}

In this paper we have obtained six novel complex PT-invariant periodic 
solutions 
of the celebrated $\phi^4$ model and hence of the focusing and the 
defocusing mKdV.
Three of these are with PT-eigenvalue +$1$ and three with PT-eigenvalue $-1$. 
These are over and above the two already known complex PT-invariant periodic 
pulse solutions with PT-eigenvalue +$1$ and two already known complex 
PT-invariant periodic kink solutions with PT-eigenvalue $-1$. Moreover, we 
have shown that there are novel connections between the complex solutions of 
several different models like DSG \cite{dsg} and DSHG \cite{razavi,khs,hks}, SG 
\cite{dj,asegur} and SHG \cite{shg} as well as between the same models like 
$\phi^4$, $\phi^2- \phi^{2n+2}-\phi^{4n+2}$ but with different values of some of 
the parameters. Using the connections we have obtained new complex PT-invariant 
periodic solutions of some of these nonlinear equations. Finally, we have also obtained 
a large number of new complex periodic solutions of a coupled $\phi^4$ model.

This work raises a number of important questions. For example, just as some of the 
complex periodic solutions simply follow from the well known real periodic solutions by 
translation, is there a similar reason presumably based on some symmetry for the 
existence of the new periodic solutions that we have obtained? If not, can one understand 
the origin of the newly discovered complex PT-invariant periodic solutions? Secondly, how 
many of these, if any, are stable? One would hope that there are certain physical situations 
where some of these solutions are likely to be relevant. In order to get a better understanding 
of some of these points, perhaps it would be worthwhile to explore similar solutions in other 
models such as SG, DSG, $\phi^6$, etc. and hence also obtain complex solutions of SHG, 
DSHG and $\phi^6$ with different values of couplings. We hope to address some of these 
issues in the near future.

\section{Acknowledgment}  A.K. is grateful to Indian National Science Academy (INSA) for 
the award of INSA Senior Scientist position at Savitribai Phule Pune University. This work 
was supported in part by the U.S. Department of Energy.


\begin{thebibliography}{99}

\bibitem{ben} C.M. Bender, Rep. Prog. Phys. {\bf 70} (2007) 947 and references theirin.

\bibitem{sch} J. Schindler, Z. Lin, J.M. Lee, F.M. Ellis and T. Kottos, 
Phys. Rev. A {\bf 84} (2011) 040101; J. Schindler, Z. Lin, J.M. Lee, 
H. Ramezani, F.M. Ellis and T. Kottos, J. Phys. A {\bf 45} (2012) 444029.`

\bibitem{ben1} C.M. Bender, B. Berntson, D. Parker and E. Samuel, Am. J. 
Phys. {\bf 81} (2013) 173.

\bibitem{pen} B. Peng, S.K. \"Ozdemir, F. Lei, F. Monifi, M. Gianfreda, 
G.L. Long, S. Fan, F. Nori, C.M. Bender and L. Young, Nat. Phys. {\bf 10}
(2014) 394.

\bibitem{rut} C.E. R\"uter, K.G. Makris, R. El-Ganainy, D.N. Christodoulides, 
M. Segev and D. Kip, Nat. Phys. {\bf 6} (2010) 192.  

\bibitem{ks1} A. Khare and A. Saxena, Phys. Lett. A {\bf 380} (2016) 856; arXiv: 1509.02899 (2015). 

\bibitem{ks2} A. Khare and A. Saxena, J. Phys. A {\bf 51} (2018) 175202.

\bibitem{as} See, e.g., M. Abromowitz and I.A. Stegun, Handbook
of Mathematical Functions (Dover Publications, New York, 2010).

\bibitem{fri} J. Cen and A. Fring, J. Phys. A {\bf 49} (2016) 365202.

\bibitem{dj} See for example, P.G. Drazin and R.S. Johnson, Solitons:
An Introduction, Cambridge Univ. Press, 1989 and references therein. 

\bibitem{asegur} M. J. Ablowitz and H. Segur, Solitons and the Inverse Scattering Transform, 
SIAM, Philadelphia, 1981. 

\bibitem{shg} M. J. Ablowitz, D. J. Kaup, A. C. Newell and H. Segur, Stud. Appl. Math. {\bf 53} (1974) 249. 

\bibitem{dsg} M. J. Ablowitz, M. D. Kruskal and J. F. Ladik, SIAM J. Appl. Math. {\bf 36} (1979) 428. 

\bibitem{razavi} M. Razavi, Phys. Lett. A {\bf 72} (1979) 89; Am. J. Phys. {\bf 48} (1980) 285. 

\bibitem{khs} A. Khare, S. Habib and A. Saxena, Phys. Rev. Lett. {\bf 79} (1997) 3797. 

\bibitem{hks} S. Habib, A. Khare and A. Saxena, Physica D {\bf 123} (1998) 341. 

\bibitem{ks3} A. Khare and A. Saxena, J. Math. Phys. {\bf 47} (2006) 092902.

\bibitem{phi4} S. Aubry,  J. Chem. Phys. {\bf 64}, 3392 (1976). 



\end{thebibliography}
\end{document}